\documentclass[10pt,amssymb,twocolumn,notitlepage]{article}

\usepackage{epigraph}
\usepackage[affil-it]{authblk}
\usepackage{dcolumn}
\usepackage[stable]{footmisc}
\usepackage{blkarray}
\usepackage{bm}
\usepackage[mathscr]{euscript}
\usepackage[font=scriptsize]{caption}
\usepackage{subcaption,graphicx}
\usepackage[table]{xcolor}
\usepackage[margin=1.7cm]{geometry}
\usepackage{tcolorbox}
\usepackage{hyperref}
\usepackage{gensymb}
\usepackage[font=footnotesize]{caption}
\usepackage{hyperref}
\hypersetup{
    colorlinks=true,
    linkcolor=blue,
    filecolor=magenta,      
    urlcolor=blue,
}
\usepackage{steinmetz}
\usepackage{amssymb}
\usepackage{makecell}
\usepackage{algorithm,lipsum,xcolor,caption}
\usepackage{enumitem}
\usepackage{amsmath,esint}
\usepackage{fancyvrb}
\usepackage{xcolor}
\usepackage{arydshln}
\usepackage{mathtools}

\definecolor{newblue}{rgb}{0.0, 0.28, 0.67}
\definecolor{newgreen}{rgb}{0.13, 0.55, 0.13}
\definecolor{newred}{rgb}{0.87, 0.72, 0.53}

\newcommand{\lbar}{\{\kern-0.5ex|}
\newcommand{\rbar}{|\kern-0.5ex\}}
\usepackage{stackengine}

\definecolor{newblue}{rgb}{0.0, 0.28, 0.67}
\definecolor{newgreen}{rgb}{0.13, 0.55, 0.13}
\definecolor{newred}{rgb}{0.87, 0.72, 0.53}

\title{Coincidence Complex Networks}
\author{Luciano da Fontoura Costa \\ \emph{luciano@ifsc.usp.br}}
\affil{S\~ao Carlos Institute of Physics -- DFCM/USP} 
\date{27rd Oct 2021}

\begin{document}

\twocolumn[
\begin{@twocolumnfalse}
    \maketitle
    \begin{abstract}
Complex networks, which are the main subject of  network science, have been wide and extensively adopted for representing, characterizing, and modeling an ample range of structures and phenomena from both theoretical and applied perspectives.  The present work describes the application of the recently introduced real-valued Jaccard and coincidence similarity indices for building complex networks from datasets. More specifically, two nodes are linked whenever their similarity is greater than a given threshold.  Weighted networks can also be obtained by taking the similarity indices as weights.  It is shown that the proposed approach allows substantially enhanced performance when compared to cosine distance-based approaches, yielding a detailed description of the specific patterns of connectivity between the nodes.   The impressive ability of the proposed methodology to emphasize the modular structure of networks is also illustrated  with respect to the iris dataset and C.~elegans neuronal network connectivity data with remarkable  results. The reported methodology and results pave the way to a significant number of possible theoretical and applied developments. 
    \end{abstract}
\end{@twocolumnfalse} \bigskip
]

\setlength{\epigraphwidth}{.49\textwidth}
\epigraph{`Invisibly linked, the universe realizes its complexity.'}
{\emph{LdaFC}}

\section{Introduction}

Though relatively recent, the area of network science (e.g.~\cite{netwsci,networks:2010,surv_meas,surv_appl}) 
has established itself not only from the perspective of theoretical contributions, but
also as a consequence of its ample range of applications to the most diverse areas
and problems~\cite{surv_appl}.   Basically, a complex network is a graph presenting structure
significantly distinct from simpler, more regular, counterparts as far as a relatively
comprehensive set of respective topological characteristics are concerned~\cite{Costa:CDT2}.

The ability of complex networks, which constitutes network science's main object of study, 
to represent virtually every discrete systems has contributed significantly to the success of this area.
This has allowed, among many other results, complex networks to be employed to 
represent several complex structures including but by no means being limited to the Internet and WWW, transportation systems including highways and airports, the structure of cities, protein interaction, 
social and scientific networks, as well as signals and images.

The representation of a structure or system as a complex network demands the adoption of
some approach for defining the network nodes, and then the application of some objective
criterion for linking these nodes.  Actually, there seems to be intrinsic complementary relationships
between connectivity, similarities and features~\cite{CostaPhysRep}.

Given that the linking of two nodes is typically expected to reflect their
shared properties, similarity-based criteria for deriving respective networks represent
a suitable choice.  Often employed approaches include the consideration of some similarity
measurement such as the cosine similarity, and to a lesser extent the adoption of set-based
alternatives, such as the Jaccard index, which has been employed almost exclusively
in cases involving binary or categorical data.

Indeed, the choice of the similarity measurement to be employed depends intrinsically on the 
nature of the data respective to the system to be represented.  In case of categorical data,
approaches such as the Jaccard index provide an interesting choice.  On the other hand,
in case the data involves integer or real-valued quantities, numeric indices such as the
cosine similarity have been frequently employed.

More recently, the Jaccard index has been generalized to real-valued 
data~\cite{CostaJaccard,CostaMset,CostaGenMops} through a multiset-based approach 
(e.g.~\cite{Hein,Knuth,Blizard,Blizard2,Thangavelu,Singh}).  Though
a version of the Jaccard similarity index adapted for multisets has been considered for
some time, it was not able to take into account the more general data taking
real, possibly negative data values.  This further generalized version of the Jaccard index 
has been called the \emph{real-valued Jaccard } similarity index.  Observe that the
ability to operate on negative values is often required in cases where the data has been
normalized, e.g.~through standardization.

In addition, the classic Jaccard index has been shown not to be
able to take into account the \emph{relative interiority} of the two compared sets, which motivated
the proposal~\cite{CostaJaccard,CostaSimilarity}  of the \emph{coincidence index}, 
which has been understood as the product of the interiority (or homogeneity) and 
real-valued Jaccard indices.  

Recent results~\cite{CostaJaccard,CostaSimilarity,CostaComparing} have indicated that product-based
similarity indices, which include both the cosine similarity and Pearson correlation coefficient,
tend to be relatively more tolerant regarding similarity quantification than non-bilinear indices based on 
combination of the minimum and maximum operations, such as the Jaccard index.  In
other words, bilinear product based similarity approaches tend to yield relatively high values for
visibly little correlated data~\cite{CostaJaccard}.  Actually, it has been 
shown~\cite{CostaSimilarity} that all these similarity approaches are ultimately related to the
Kronecker's delta function, which can be understood as the prototypical approach to quantifying
the similarity between two values, though in a quite strict manner.  

Therefore, the real-valued  Jaccard and coincidence indices, both of which relying on 
maximum and minimum operations, constitute interesting alternatives to be considered
in situations where the cosine, Pearson correlation, and classic Jaccard have been
traditionally adopted.  In fact, it has been shown~\cite{CostaComparing,CostaIndNeur} that 
the real-valued Jaccard, and in  particular the coincidence indices tend to allow substantially 
more detailed and selective similarity quantification, being capable of removing secondary, 
smaller spatial scale noise and structure while emphasizing the coincidence matches in the form 
of sharp, narrow matching peaks.

Another problem in network science that has been as important, or perhaps more important,
than the critical task of transforming datasets into respective networks concerns the
issue of, given a network, how to identify its respective modular structure, or 
communities (e.g.~\cite{networks:2010}).  The particular importance of this problem stems from the
fact that several real-world networks of special interest --- including the Internet, WWW,
and social networks --- present a modular organization that is critically important not
only for better understanding the structures of these systems, but also because it is known
that  the network modularity is is capable of strongly influencing several types of dynamics
taking place on the respective networks.    Though several methods have been proposed
for community finding, this problem remains to a great extent open because of the
intrinsic difficulties involved.  In fact, as with clustering in pattern recognition, the
identification of communities that are not originally well-separated represents a
particular challenge for the modularity methods.  Remarkably, the real-valued
Jaccard and coincidence indices-based approach reported in the present work
is shown to allow an impressive effect in enhancing the modular structure of existing
networks.

The present work aims at studying the potential of the real-valued Jaccard and
coincidence similarity indices as the basis for mapping datasets, involving one or more
features corresponding to real values, into respective complex networks.  The
possibility to start with a weighted or unweighted network and then obtain features
for each node corresponding to the respective topological properties is also 
investigated.  This network can then be transformed by taking into account the
the similarity indices between the respectively obtained topological measurements.

It is shown not only that the proposed method allows an impressive level of
detail about the specific interconnecting structure of datasets when represented
as suggested here, but also a remarkable ability to emphasizing the network modularity.   
When applied as a means to transform an existing network into a new representation
based on some of the topological property of the nodes, the proposed method
was able to emphasize significantly the respective modular structure to the point that
even the simplest community finding approaches will succeed in identifying the respective
communities.

In order to illustrate the potential of the proposed approach for obtaining detailed information 
about the data element interrelationships, as well as on the overall data modularity, we
apply the representation method to the iris dataset~\cite{IrisDataset}, 
calculating the real-valued Jaccard and coincidence indices between a standardized
version of the four original features, with remarkable results.  Several other examples of
application of the coincidence methodology for transforming generic datasets into respective
networks can be found at~\cite{CostaCaleidoscope}.

In addition, we illustrate
that the proposed methodology also provides an effective means for enhancing the
modular structure of already existing networks.  This possibility is illustrated
with respect to the \emph{C. elegans} data, with impressive results.  Actually, these
results indicate that the proposed methods not only provide an effective and detailed
representation of the data interrelationships, but can also be applied as a valuable
resource for  community finding respectively to any network.

Though undirected complex networks are henceforth assumed in this work, the
reported concepts and methods can be readily extended to directed complex
network research.

\section{Basic Concepts}

A complex network is basically a graph, being therefore composed of $N$ nodes or vertices
and $E$ connections, edges, or links.  Complex networks are often represented in terms
of the respective adjacency matrices or lists of edges.  The former of these possible
representations involves using a matrix $A$ where a connection from a node $j$ to
a node $i$ implies $A\left[i, j\right]=1$, with $A\left[i,j\right]=0$ being otherwise enforced.

Given a network represented in terms of its respective adjacency matrix (or other
representations as an adjacency list), several respective
topological measurements (e.g.~\cite{surv_meas}) can be obtained, some of which
are adopted in the present work and briefly reviewed as follows (e.g.~\cite{surv_meas}).

The \emph{degree} of a node $i$ corresponds to the respective number of connections
established with that node.  It can be readily calculated as:
\begin{equation}
   k_i = \sum_{j=1}^N A\left[i,j\right]
\end{equation}

The \emph{clustering coefficient} or \emph{transitivity} of a node provides a quantification
of how well the neighbors of a node are interconnected, corresponding to real values
varying between 0 and 1.

The \emph{eccentricity} of a network node corresponds to its shortest distance between
that node and the farthest other node in the same network.

The \emph{page rank} (e.g.~\cite{pagerank}), used in WWW browsers (Google), this index 
applies to each network node, but needs to be obtained while taking all the nodes into account.
This measurement is related to the principal eigenvalue of the network matrix. 

Possibly the most frequently adopted means for quantifying the similarity between
network nodes consists of  the \emph{cosine similarity}, which relies on the
inner product between two sets $\vec{x}$ and $\vec{y}$, being expressed as:
\begin{equation}
   \left< \vec{x}, \vec{y} \right> =  \sum_{i=1}^N x_i \ y_i =  |\vec{x}| |\vec{y}| \cos(\theta) 
   \Longrightarrow    cos(\theta) =  \frac{ \left< \vec{x}, \vec{y} \right>}{\left< \vec{x}, \vec{y} \right>}  
\end{equation}

In order to avoid biases typically implied by varying magnitudes between the several features,
the statistic linear transformation of \emph{standardization} is frequently adopted prior to
similarity quantification. This transformation can be performed as follows:
\begin{equation}
   \tilde{f}_i = \frac{f_i - \mu_i}{ \sigma_i}
\end{equation}

It can be shown that each of the resulting features will have zero means and unit standard
deviation.  In addition, most of the feature values become comprised within the interval $\left[-2,2\right]$,
though this depends o their respective statistical distributions.  Conveniently, the
newly obtained variables also become non-dimensional.

Observe that the application of the standardization procedure intrinsically implies that
any index or measurement applied to quantify the similarity between data elements
will need to be able to cope with negative values, which is not the case of the classic
Jaccard index, or even its standard multiset generalization to non-negative values.

\section{Real-Valued Jaccard and Coincidence Similarity Indices}

The classic Jaccard index~\cite{Jaccard1,jac:wiki,CostaJaccard}) for quantifying the similarity between
two sets $A$ and $B$ can be defined as follows:
\begin{equation}
   J(A,B) = \frac{|A \cap B|}  {|A \cup B|}
\end{equation}

where $|A|$ is the cardinality (an non-negative integer number) of set $A$.

A recent generalization~\cite{CostaJaccard,CostaSimilarity,CostaGenMops}, henceforth referred to as 
\emph{real-valued Jaccard similarity index}, allows to take into account real, possibly negative data values~\cite{CostaJaccard,CostaMset,CostaSimilarity}.
The real-valued Jaccard similarity index can be expressed as:
\begin{align}  \label{eq:rvjac}
     &s_1(A,B) =   \frac{ \sum_{i\in S} s_{x_i y_i}\min\left\{s_{x_i} x_i,  s_{y_i} y_i \right\} }{\sum_{i \in S} \max\left\{ s_{x_i} x_i, s_{y_i} y_i \right\} }   
\end{align}

where $x_i$ and $y_i$ correspond to the multiplicities of the sets $A$ and $B$, $s_x = sign(x)$,
$S$ is the combined support of multisets $A$ an $B$,
$s_y = sign(y)$, $s{xy} = s_x s_y$, and $S$ is the combined support  of the multisets, 
which generally corresponds to the union of the elements in $A$ and $B$.

Observe that, as in~\cite{Akbas1}, the numerator in Equation~\ref{eq:rvjac} is capable of taking 
into  account the signs of the two values $x$ and $y$.   That numerator has also been shown
to correspond to the intersection between two real-valued multisets~\cite{CostaGenMops}:
\begin{align}  
     x \sqcap y = \sum_{i \in S}  s_{x_i,y_i} \min\left\{s_{x_i} x_i,  s_{y_i} y_i \right\} 
\end{align}

required so that $x \sqcap \Phi = \Phi$, where $\Phi$ is the empty multiset.

Another possibility for taking into account
the signs is as described in~\cite{Mirkin}:
\begin{align}  \label{eq:mirkin}
     s_+ =  \sum_{i \in S}  | s_{x_i} + s_{y_i} | \min\left\{s_{x_i} x_i,  s_{y_i} y_i \right\} /2
\end{align}

which allows only the cases where the signs are aligned (i.e.~$s_x =s_y$) to contribute to the
sum in Equation~\ref{eq:rvjac}. However, the opposite situation in which $s_x =-s_y$ can be
expressed by using the following quantity~\cite{CostaSimilarity}:
\begin{align}  \label{eq:mirkin}
     s_- =  \sum_{i \in S}  | s_{x_i} - s_{y_i} | \min\left\{s_{x_i} x_i,  s_{y_i} y_i \right\} /2
\end{align}

So that we have separated the contributions of the positive and negative sign alignments.
These two indices can then be combined as:
\begin{equation}
     s_{\pm,\alpha} =  \sum_{i \in S}  \left[ \alpha \right] s_+ -\left[ 1- \alpha \right] s_-
\end{equation}

where $0 \leq \alpha \leq 1$ is a parameter controlling the relative weights of the aligned
and anti-aligned indices $s_+$ and $s_-$.  It can be verified that, if $\alpha = 0.5$, we have
\begin{equation} 
    s_{\pm,0.5} = \sum_{i \in S} s_{x_i y_i}\min\left\{s_{x_i} x_i,  s_{y_i} y_i \right\} 
\end{equation}

Another interesting possibility is to use the index:
\begin{equation} 
     p = \frac{ \sum_{i \in S} x_i \ y_i } { \left( \sum_{i \in S} \max\left\{s_{x_i} x_i,  s_{y_i} y_i \right\} \right)^2  }
\end{equation}

which has also been verified to be identical to Equation~{eq:rvjac}.  

The classic Jaccard index, as well as its generalization to multisets and real-values,
have been shown not be able to take into account the relative interiority of the two 
sets~\cite{CostaJaccard}.  

When adapted to real values~\cite{CostaJaccard,CostaSimilarity} The relative interiority 
(or overlap e.g.~\cite{Kavitha}) index between any two multisets can be expressed
as: 
\begin{eqnarray}
   I(x,y) = \frac{\sum_{i \in S} \min\left\{s_{x_i} x_i, s_{y_i} y_i \right\}} 
   {\min\left\{\sum_{i \in S} s_{x_i} x_i , \; \sum_{i \in S} s_{y_i} y_i \right\}}  
\end{eqnarray}

This measurement, however, does not take into account the intersected region between
the two multisets, a property that is effectively reflected in the
Jaccard index.  

In oder to combine the best features of each of the above two measurements, both 
of which relating directly to the concept of similarity between two sets, the 
\emph{coincidence index} between any two multisets $A$ and $B$ has been
proposed~\cite{CostaJaccard,CostaMset} as corresponding to the product of the
respective interiority and Jaccard indices.  Therefore, we have that the coincidence
index between any two sets $A$ and $B$ can be calculated as:
\begin{equation}
   C(x,y) =  I(x,y) J(x,y) 
\end{equation}

with $0 \leq C(x,y) \leq 1$.

The coincidence index tends to provide one of the most strict and detailed quantification of
the similarity between two mathematical structures, having allowed impressive performance
respectively to pattern recognition~\cite{CostaComparing,CostaIndNeur} and hierarchical
clustering~\cite{CostaCluster}.  In the former case, the coincidence index was shown to be
able to yield sharp, narrow peaks indicating the matches between a template and an object
signal, while substantially attenuating secondary, small scale noise and otherwise unwanted
background structure.  

As such, the coincidence index can be understood as a binary operator
(in the mathematical sense of taking two arguments) that combines low and high-pass filtering
in order to best suit the pattern recognition objectives.  This critically important feature is a direct
consequence of the non-bilinearity of the operations maximum and minimum which are part of the
definition of the coincidence index.

Even more important is the fact that the normalization in Equation~\ref{eq:rvjac} yields
a similarity profile almost identical to the generalized Kronecker delta function~\cite{CostaSimilarity}:
\begin{equation}
    \delta^{\pm}_{i,j} = sign(i \ j) \ \delta_{i,j}
\end{equation}

\section{Building Coincidence Networks}

In this work, we will approach the situation in which a set of $N$ data observations, samples,
element or individuals, have been priorly characterized in terms of $M$ respective measurements
or \emph{features}.
This can be organized as a table or matrix where each of the $N$ rows corresponds to 
a data observation, while the columns contain the respective measurements or features
$x_i$, $i = 1, 2, \ldots, M$. Experimental data are normally represented and stored in this format.

Given such a table, we aim at obtaining a respective representation as a complex network.
The commonly adopted procedure consists of understanding each data element as a node,
which links are established based on the similarity (or difference) between the respective
features.

Frequently applied methodologies rely direct or indirectly on the  cosine distance between the 
involved features.    Here, we adopt the recently introduced real-valued Jaccard and coincidence 
indices.

The basic methodology proposed in this work therefore consists of: (i) standardizing the original
dataset; and (ii) obtaining the pairwise similarities by using the cosine, real-valued Jaccard, or
coincidence indices.  The adjacency matrix of the resulting network can be taken by thresholding
the respective similarity matrix (understood as a weight matrix) by a value $T$.  It is also possible to 
preserve the weights of the thresholded links, resulting in the representation of the original dataset in
terms of a respective weight network.  The latter approach, which preserves more information
about the interrelationship between the original data elements, is henceforth adopted in this work.

\section{The Iris Dataset}

For simplicity's sake, we consider the iris dataset (e.g.~\cite{IrisDataset}), which has been 
extensively considered in pattern recognition and machine learning.   This dataset contains 50 
samples of each of 3 iris flower species, namely \emph{Iris setosa}, \emph{Iris virginica} and 
\emph{Iris versicolor}.  Each of the data elements is characterized in terms of four
respective real-valued features.  All the network visualizations presented in this work
were obtained by using the Fruchterman-Reingold agorithm (e.g.~\cite{Fruchterman}).

\begin{figure}[h!]  
\begin{center}
   \includegraphics[width=0.9\linewidth]{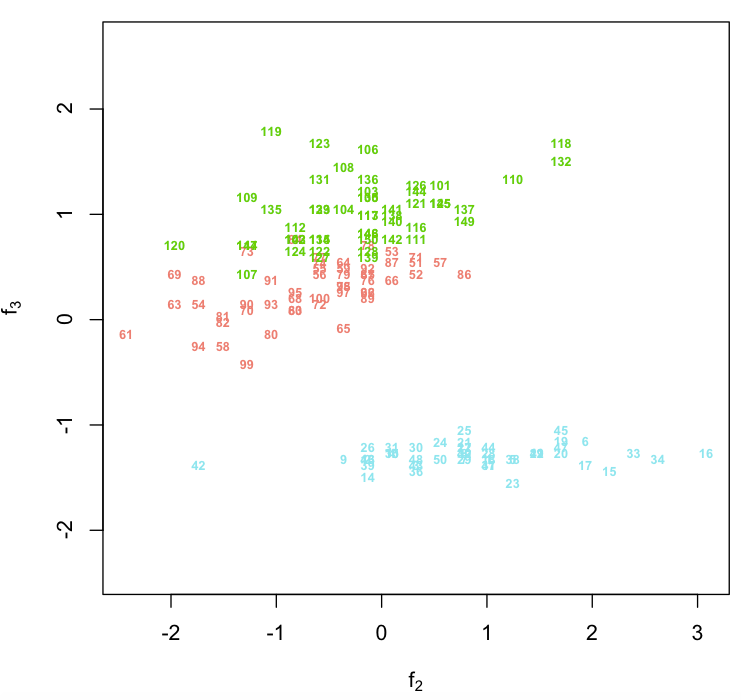}  
    \caption{Scatterplot of the Iris dataset considering original features 2 and 3.
    Each of the three iris species have been shown in specific respective colors.
    The numbering is as in the original dataset, with 50 individuals of each species
    number subsequently from 1 to 50, 51 to 100, and 101 to 150.  One group
    is markedly separated, but the other two are adjacent and even present 
    overlap.}
    \label{fig:scatter}
    \end{center}
\end{figure}

Figure~\ref{fig:scatter} depicts the iris data set considering, for visualization purposes,
only the second and third original features.  All four real-valued original features are
otherwise taken into account henceforth.  In addition, in order to avoid effects of the 
varying magnitudes of the four original measurements, which could therefore bias the 
network representations, we consider the respectively standardized dataset.  In all 
described cases, the obtained complex network is thresholded for the sake of less 
cluttered visualization, but the weight values of the remaining nodes are retained.

The overall structure of the iris dataset is, however, difficult to be observed as a whole
because it involves a four-dimensional space.  Nevertheless, we can see from Figure~\ref{fig:scatter} 
that this dataset contains a well separated cluster (in blue), while the other
two groups are adjacent and present overlap.  It has been particularly difficult to separate 
these two clusters in the literature.

\begin{figure}[h!]  
\begin{center}
   \includegraphics[width=0.9\linewidth]{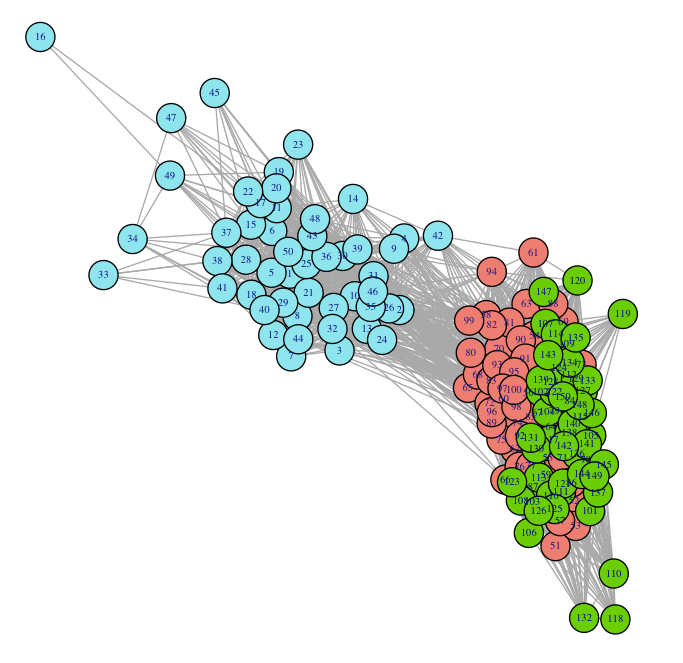}  
    \caption{Complex network of the iris dataset obtained while considering
    the \emph{cosine similarity} and all four original features.  This result mirrors
    the relative distribution of points in the original dataset, without providing
    detailed information about specific connectivity patterns.}
    \label{fig:Euclidean}
    \end{center}
\end{figure}

The application of the cosine similarity to obtain a respective complex network
resulted in the structure shown in  Figure~\ref{fig:Euclidean} .

\vspace{0.5cm}
\vspace{0.5cm}

The obtained complex network resembles the relative position and separation
of the original dataset as projected onto the second and third original measurements
shown in Figure~\ref{fig:scatter}, without providing additional information about the
specific patterns of interconnectivity.  However, a rather generic pattern of interconnectivity
is obtained that does not add much to the understanding about the more specific
and detailed interrelationship between the original data elements.

\begin{figure}[h!]  
\begin{center}
   \includegraphics[width=0.9\linewidth]{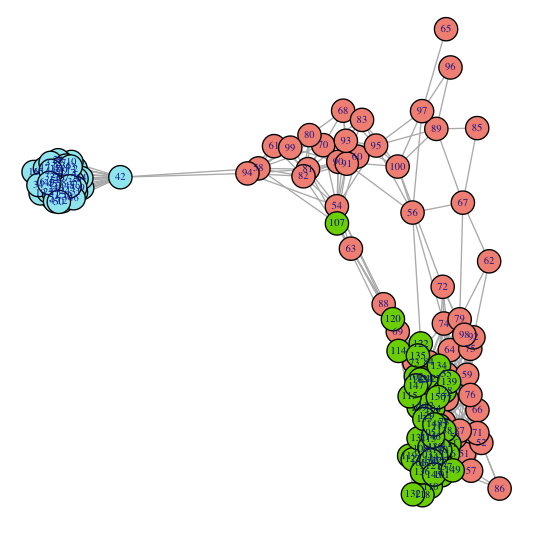}  
    \caption{Complex network of the iris dataset obtained by using the
    \emph{real-value Jaccard index} and the four original features.  Several remarkable 
    characteristics can be identified in this result.  First, we have that the blue group, 
    which was already well separated in the original feature space, coalesced into
    a very compact community.  Second,  we have that a substantial portion of the
    salmon group resulted relatively well separated from the other two groups.
    In addition, the obtained enhanced modularity indicates the improved 
    discriminability allowed by this approach.}
    \label{fig:Jaccard}
    \end{center}
\end{figure}

Now, we proceed to applying the recently introduced real-valued Jaccard index
to the iris dataset, yielding the result depicted in Figure~\ref{fig:Jaccard}.

\vspace{0.5cm}

This resulting network reveals a surprising separation between the groups.  Not only the blue group
has coalesced into a very compact community, but a substantial portion of the salmon
group has also resulted separated from the other two groups.   In addition, this
result allows a substantially more detailed and specific representation of the specific patterns of
interrelationship between the involved data elements, as well as the underlying modular 
structure.   Observe also that the outliers points in Figure~\ref{fig:scatter} have been duly 
incorporated at the border of the obtained network.

In a sense, the adoption of the real-valued Jaccard similarity acted almost as a 
community detection algorithm, strongly emphasizing the respective modular structure
and centrality of the nodes.

\begin{figure}[h!]  
\begin{center}
   \includegraphics[width=0.9\linewidth]{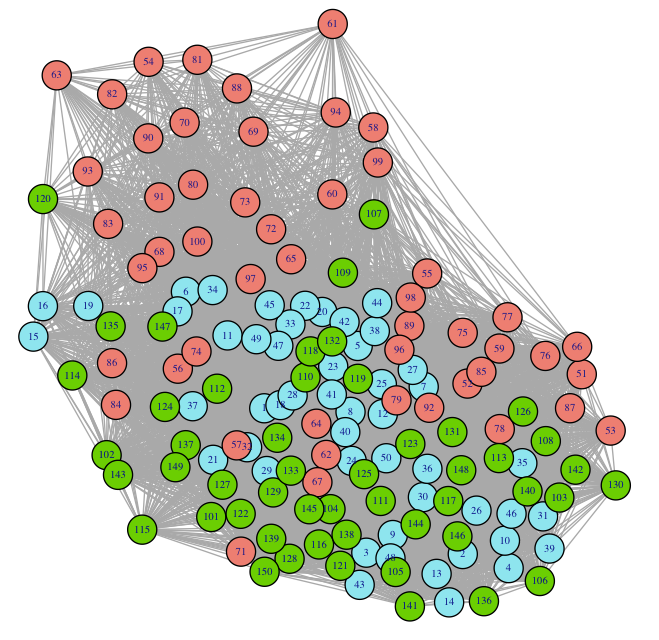}  
    \caption{Complex network of the iris dataset obtained from the \emph{interiority
    index}, which can also be understood as quantifying similarity,
     between each pair of original data elements.  The obtained
    results corroborate that, despite the ability of the interiority index to
    quantify the relative interiority between two multisets, this index does not actually
    take into account the relative intersection are with respect to the
    size of the msets, therefore being not able to reveal much about the
    specific interconnectivity patterns and modularity.  Observe that, by multiset, we are meaning 
    the multiset of 
    features associated to each data individual.}
    \label{fig:Interiority}
    \end{center}
\end{figure}

In order to illustrate the interiority similarity,  we obtained the respective complex
network shown in Figure~\ref{fig:Interiority}.

\vspace{0.5cm}

This result well illustrates that, despite the effectivity of the interiority index in
quantifying the relative interiority of a pair of sets, this index by itself cannot actually
provide a more complete picture of the similarity between the several involved
data elements.

\begin{figure*}[h!]  
\begin{center}
   \includegraphics[width=0.9\linewidth]{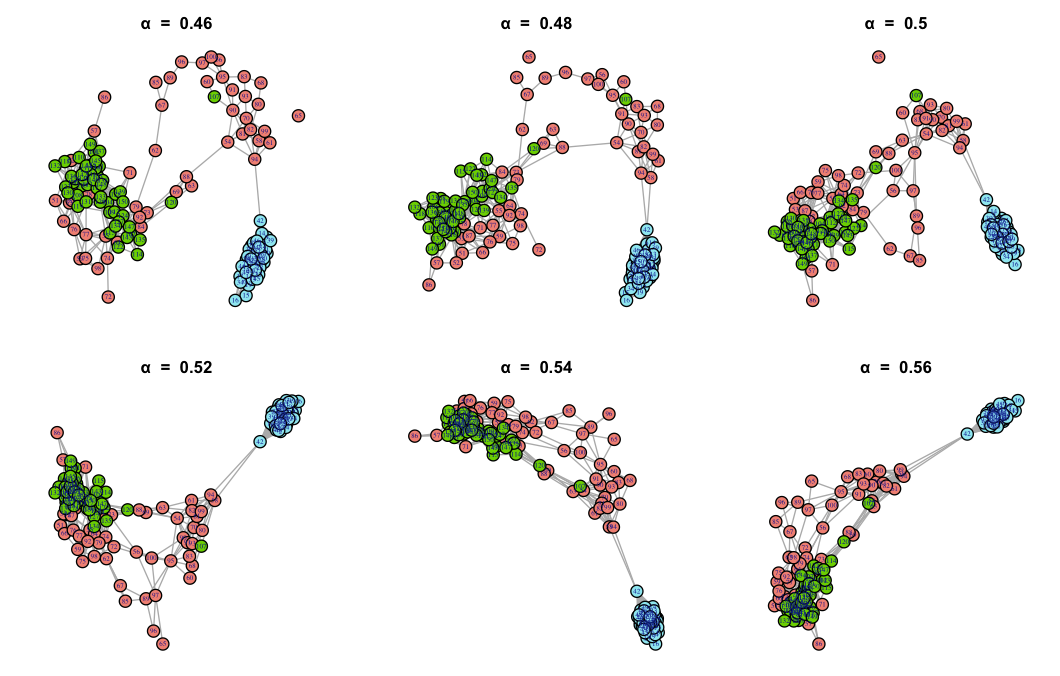}  
    \caption{Complex network of the iris dataset obtained by using the
    \emph{coincidence index} on the four original features for 
    $\alpha=0.46, 0.48, 0.5, 0.52, 0.54, 0.56$ and threshold $T = 0.5$.
    The parameter $\alpha$ allowed the control of the effect of the 
    opposite sign pairs of features, with the most separated structures
    being obtained fro $\alpha=0.46$, in which case not only 
    the  blue cluster resulted well-separated (as in the other cases), but
     a good separation was also observed between a good portion of the
     other two clusters. Even the simplest community finding algorithm
     will be able to separate the resulting network into 3 or 4 modules.}
    \label{fig:Coincidence}
    \end{center}
\end{figure*}

Figure~\ref{fig:Coincidence} shows the network obtained by adopting the
coincidence index with $alpha=0.46, 0.48, 0.5, 0.52, 0.54, 0.56$ and threshold $T = 0.5$.

\vspace{0.5cm}

The already identified enhanced discriminative potential of the coincidence index~\cite{CostaMset,CostaComparing} has allowed a network representation of the
iris dataset that is even more detailed and structured regarding the interrelationship
between the data elements, as well as the overall data modular structure.  
A particularly detailed network has obtained respectively to $\alpha=0.46$, 
in which not only the blue cluster resulted significantly separated from the others,
but some separation can be observed also regarding the green and salmon groups.

\begin{figure}[h!]  
\begin{center}
   \includegraphics[width=0.89\linewidth]{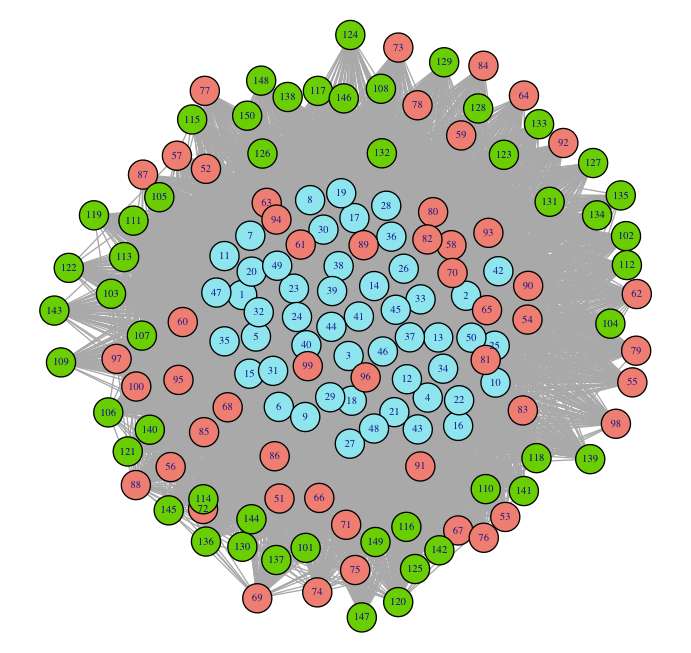}  
    \caption{Network obtained for negative coincidence values.  The connections
    are therefore established between data elements that present mutual negative
    similarity, indicating anti-coincidences.  Recall that the four original measurements
    have been standardized, therefore becoming mostly distributed between -2 and 2.}
    \label{fig:neg_coincidence}
    \end{center}
\end{figure}

GIven that both the real-valued Jaccard and coincidence indices convey information
about the sign of the joint variation of the two multisets, it becomes possible to derive
respective complex networks considering only the negative coincidence similarities.
In the case of the iris dataset, the obtained network considering only negative
coincidence values (i.e.~$\alpha=0$) is shown in Figure~\ref{fig:neg_coincidence}.

Interestingly, two hierarchical levels, or modules, have been obtained. The
most central one corresponds to the blue group.    The other two groups, which
are more intertwined and dispersed, resulted in the outer community.  
The outliers in the original dataset resulted
at the border of the obtained network, while the most centra data elements are to be 
found at the core of the obtained network.  This indicates that the negative coincidence
network provides an effective method for taking into account the centrality, and outliers, 
of the data elements.

\section{Community Finding in the \emph{C. elegans} Network}

\begin{figure}[h!]  
\begin{center}
   \includegraphics[width=0.89\linewidth]{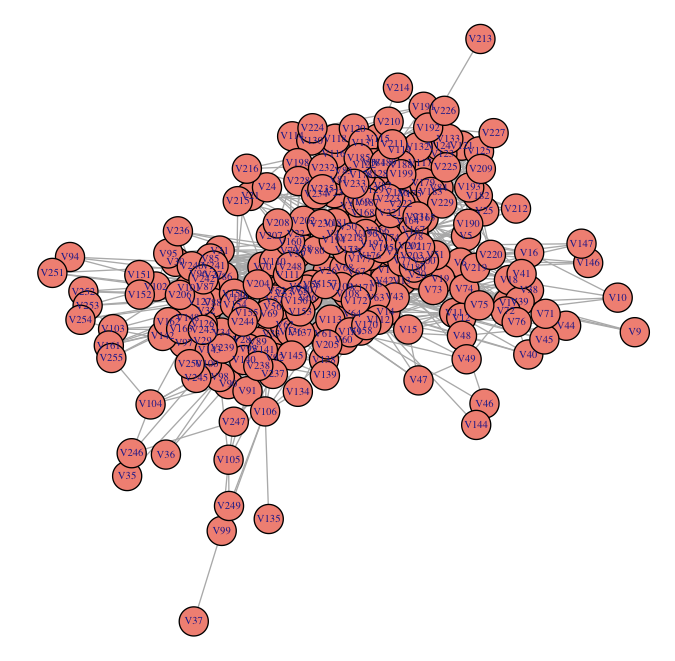}  
    \caption{Visualization of the \emph{C. elegans} considering its
    original unweighted adjacency matrix.}
    \label{fig:Celegans_input}
    \end{center}
\end{figure}

Figure~\ref{fig:Celegans_input} illustrates the \emph{C. elegans} network~\cite{Celegans}
as visualized from its respective \emph{unweighted} adjacency matrix.  

\vspace{0.5cm}
\vspace{0.5cm}

Instead of representing the network from the data elements features, as was
done in the case of the iris dataset, now we apply the proposed representation
methodology on four estimated topological measurements
of the original network (e.g.~\cite{surv_meas}) --- namely degree, transitivity, page rank, and eccentricity ---
and considered these measurements as respective features of each node (e.g.~\cite{CommHuang}).

\begin{figure*}[h!]  
\begin{center}
   \includegraphics[width=0.9\linewidth]{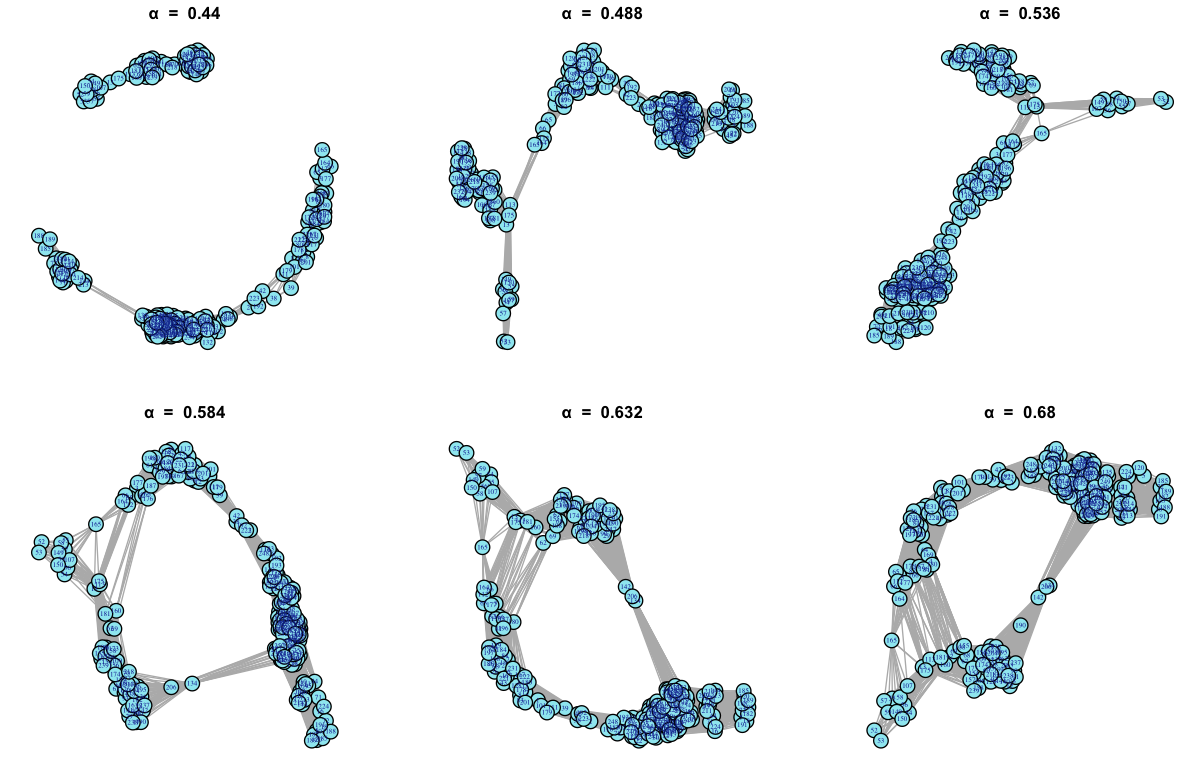}  
    \caption{The network representation of the \emph{C. elegans} data
    obtained from only four topological features calculated for each node
    while taking into account the respective unweighted adjacency matrix,
    respectively to for $\alpha = 0.44, 0.488, 0.536, 0.584, 0.632, 0.68$
    and $T=0.5$. 
    All the obtained results emphasize the known modular structure of this network,
    while providing additional detailed information about nodes interrelationships.}
    \label{fig:Celegans_coinc}
    \end{center}
\end{figure*}

The resulting representation obtained by using the coincidence index is depicted
in Figure~\ref{fig:Celegans_coinc} for $\alpha = 0.44, 0.488, 0.536, 0.584, 0.632, 0.68$
and $T=0.5$.    Remarkably, well-defined modular structures have been obtained throughout, 
without need of any community finding methodology.  It is of particular interest to observe
the effect of the parameter $\alpha$ in controlling the overall modularity degree.
This examples well-illustrates the  potential of the proposed methodology as a simple and 
effective means for emphasizing and visualliy identifying communities while taking into
account only a handful of topological properties of each of the involved nodes.

\vspace{0.5cm}

\section{Concluding Remarks}

Network science has consolidated itself as an important area, with many theoretical
and applied contributions.  Much of this success stems
from the ability of complex networks to \emph{represent} virtually every possible discrete
structure or system, while continuous counterparts can always be discretized to
a given resolution.  

One of the enduring challenges in network science concerns how to transform datasets
into networks, which often constitutes the very first step while analyzing several types
of data.   Though product-based measurements such as the cosine similarity have been 
ubiquitously applied, other measurements based such as the Jaccard index
have  been mostly constrained to categorical or binary data.

The present work set out at addressing the possibility of using the recently introduced
real-valued Jaccard and coincidence indices as the basis for deriving complex network
representations of real-valued data, while also encompassing other types of data
such as integer and categorical.  Actually, an immediate benefit of such an approach
constitutes in its being inherently capable of taking into account hybrid datasets
involving several types of features, which is allowed by the fact that both the real-valued
Jaccard and coincidence index can immediately adapt to these other types of hybrid data.

The results obtained with respect to the well-known iris dataset revealed some remarkable
findings.  First, we have that the enhanced precision and discrimination potential of the
real-valued and Jaccard indices, already observed experimental~\cite{CostaComparing}
and theoretically~\cite{CostaSimilarity}, allowed complex networks to be obtained which
reveal to a great level of detail the intrinsic relationships between the data elements as
well as the overall data modular structure.  In other words, the maximum and minimum
based similarities (two non-bilinear operations directly related to multisets), provide a markedly
more specific and detailed description of the data.   Several additional examples of
datasets visualized as networks by using the proposed methodology can be found 
at~\cite{CostaCaleidoscope}.

In order to illustrate the impressive performance of the proposed methodology for
enhancing the modular structure of networks, we applied it to the \emph{C. elegans}
unweighted network.  In this case, instead of calculating the similarity indices from the features
associated to each data element, we obtained a set of only four topological measurements
of the network in order to characterize the connectivity of each node.  All resulting
networks, obtained for several values of the parameter $\alpha$  were markedly modular, 
adhering surprisingly well with the overall known  structure of this model animal.

Comparatively, the cosine similarity-based
obtained representation can be understood as a direct translation of the 
original pairwise distances between the original datasets, without revealing any additional
structure or properties of the data elements interrelationship.
In a sense, the real-valued Jaccard and coincidence indices action resulted effectively
in a particularly useful resource for the particularly challenging problem of
community detection.  Observe that  the choice of the similarity measurement, as well
as the frequently adopted standardization will have respective effects on the obtained
results.  The adoption of these approaches therefore
needs to take into account demands and requirements specific to each application.
It is also important to recall that data analysis results should be always understood only
as hypotheses to be further verified and validated.

All in all, the interesting features provided by the reported methodology include: (i) substantial
effectiveness for revealing the pattern of interconnections between the original data
elements; (ii) impressive potential for enhancing the modular structure of the original data;
(iii) can be applied on datasets involving real-valued, integer, categorical and binary types: 
(iv) can be applied on topological features derived from the adjacency matrix of data
already in network format: (v) is based on non-bilinear operations (including the maximum
and minimum) that can substantially attenuated noise and secondary structures (low-pass
filtering) simultaneously as the more meaningful structures are enhanced (high-pass
filter), an effect that cannot be achieved through the most commonly used bilinear
operations such as correlation and cosine similarity: (vi) remarkable conceptual and
computational simplicity, relying on the maximum, minimum, sign and
division operations; (vii) employs an improved version of the well-known and extensively
used (for categorical data) Jaccard index; (viii) incorporates additional verification of
the relative interiority between the data;  (ix) is founded on a formal framework corresponding
to the generalization of multisets to real, possibly negative values; (x) incorporates a
parameter $\alpha$ allowing the separated control of the contributions of the positive
and negative combinations of sign alignments in the original data; (xi) does not
involve any matrix operation such as inverse, pseudo-inverse, or even products; 
(xii) the original features (measurements) remain in their original domain, except for the
eventual standardization, not being mapped into other less intuitive spaces; (xiii) can be
readily adapted to take into account similarity between three or more data elements~\cite{CostaJaccard};
and (xiv) the implemented similarity comparisons can have a probabilistic interpretation given
the Jaccard construction and normalization.
 
The proposed methodology bears significant potential for ample application in complex
network research, allowing a unified and effective manner of obtaining representations
of the most diverse type of data and features, including cases of mixed types of
measurements.    Another interesting possibility consists in adopting the other
similarity indices proposed in~\cite{CostaJaccard}, especially the addition-based
multiset Jaccard for real-valued data.

\vspace{0.7cm}
\emph{Acknowledgments.}

Luciano da F. Costa
thanks CNPq (grant no.~307085/2018-0) and FAPESP (grant 15/22308-2).  
\vspace{1cm}

\bibliography{mybib}
\bibliographystyle{unsrt}

\end{document}